# BETTER BOUNDS FOR FREQUENCY MOMENTS
# IN RANDOM-ORDER STREAMS

ALEXANDR ANDONI, ANDREW MCGREGOR, KRZYSZTOF ONAK, AND RINA PANIGRAHY

ABSTRACT. Estimating frequency moments of data streams is a very well studied problem [1–3,9,12] and tight bounds are known on the amount of space that is necessary and sufficient when the stream is adversarially ordered. Recently, motivated by various practical considerations and applications in learning and statistics, there has been growing interest into studying streams that are randomly ordered [3,4,6–8,11]. In the paper we improve the previous lower bounds on the space required to estimate the frequency moments of a randomly ordered streams.

## 1. INTRODUCTION

Consider a stream $\langle a_1, \ldots, a_m \rangle$ where each $a_i \in [n]$. The $k$-th frequency moment is defined as

$$F_k = \sum_{i \in [n]} f_i^k$$

where $f_i = |\{j : a_j = i\}|$. It is known that $\tilde{\Theta}_\epsilon(n^{1-2/k})$ space is necessary and sufficient to estimate $F_k$ in the data-stream model when the stream is ordered adversarially [5, 9]. Recently, there has been a growing interest in understanding the data-stream model when the stream is determined by a set of $m$ elements and a random permutation of these elements [3, 4, 6–8, 11]. Here the goal is to understand the amount of space and/or passes that is required to solve a problem with large probability where the probability is taken over both the coin flips of the algorithm and the random permutation of the stream. For some problems, significantly less resources are required in this model, e.g., it was shown that any $O(\text{polylog } n)$-space algorithm for finding the median of a length-$n$ stream with $9/10$ probability requires $\Omega(\log n / \log \log n)$ passes in the adversarial-order model whereas in the random-order model, $O(\log \log n)$ passes suffices. For further details and a motivation for the random-order model, including its relevance to applications in learning and statistics, see [8, 10].

1.1. **Previous Result and Our Result.** The previous best lower bound for estimating $F_k$ in a random-order data-stream model was $\Omega(n^{1-3/k}/\log n)$. The hardness instance (based on the unique intersection promise in the multi-party set-disjointness problem) consisted of either a) at most $n$ elements of multiplicity one or b) $\Omega(n)$ elements of multiplicity one and an element of multiplicity $n^{1/k}$. The bound follows by considering the communication when the elements are partitioned uniformly at random between $P = \Theta(n^{2/k})$ players. With high probability over the random partition, it can be shown that any one-way protocol requires sending of $O(n^{1-1/k})$ bits in total and hence at least one message must require $\Omega(n^{1-3/k})$ bits of communication in a constant round protocol [3]. This yields a $\Omega(n^{1-3/k}/\log n)$ space lower bound for the single pass data-stream problem that is based on the observation that if the $i$-th player randomly orders their elements to form a stream $s_i$, then the stream formed my concatenating these streams $\langle s_1, s_2, \ldots, s_P \rangle$ is in random order.

Note that the above communication bound is tight in the sense that if $P$ was $o(n^{2/k})$ then, with large probability in case b) at least one player would receive two identical elements while in case a) this would can not happen since there are no duplicate elements. Our new approach sidesteps this issues and we prove that $\Omega(n^{1-2.5/k})$ bits of space is required to estimate $F_k$. The best upper-bound known for estimating $F_k$ in random-order streams is the same as in the case of adversarial streams, i.e., $\tilde{\Theta}_\epsilon(n^{1-2/k})$. We conjecture that the actual space complexity for $F_k$ in the random-order model is $\tilde{\Theta}_\epsilon(n^{1-2/k})$. In other words, that frequency moments, unlike the median problem, is just as hard in the random-order model as it is in the adversarial-order model.

## 2. The New Bound

At the heart of our proof is a reduction from $t$-party set-disjointness. An instance of this problem consists of $t$ subsets $S_i \subset [N]$ where the $i$-th player knows only $S_i$. These subsets satisfy the condition that each $j \in [N]$ appears in either $0, 1$, or $t$ of the subsets. The problem is to determine if there exists $j$ such that $j \in S_i$ for all $i \in [t]$. Furthermore, we may assume that $|S_1| = |S_2| = \ldots = |S_t| = cN/t$ for some arbitrarily small constant $c > 0$. It was shown that any randomized protocol (maybe using public random bits) that solves $t$-party set-disjointness with probability $2/3$ requires $\Omega(N/(t \log N))$ bits of communication [5].

Our argument works by assuming the existence of an $s$-space, single-pass, data-stream algorithm that returns a 2-approximation for $F_k$ of a $O(n)$-length stream with probability $99/100$ on the assumption that the order of the stream elements is chosen uniformly at random from the set of all orderings. We use this algorithm to construct a communication protocol for $t$-party set-disjointness when $N = O(n^{1-1/(2k)})$ and $t = \Omega(n^{1/k})$. The protocol uses $O(sn^{1/k})$ bits and we therefore deduce that $s = \Omega(n^{1-2.5/k}/(\log n))$.

Before we present this protocol, we present two preliminary lemmas that will be important.

**Lemma 1.** *Let $\mathcal{I} = \{I_1, \ldots, I_t\}$ be $t = n^{1/k}$ random sets from $\text{Cycle}_{n,w} := \{\{i - 1 \ (\text{mod } n) + 1, \ldots, w + i - 2 \ (\text{mod } n) + 1\} : i \in [n]\}$ where $w = c_1 n^{1-3/(2k)}$. For small enough $c_1$, with probability at least $99/100$,*
  *(1) $I_{i_1} \cap I_{i_2} \cap I_{i_3} = \emptyset$ for any $i_1 < i_2 < i_3$.*
  *(2) $|\{(i_1, i_2) : i_1 < i_2, I_{i_1} \cap I_{i_2} \neq \emptyset\}| \leq n^{1/(2k)}$.*

*Proof.* We may assume that $w|n$ by adjusting $c_1$ and we partition $[n] = J_1 \cup \ldots \cup J_{n/w}$ where $J_i = \{1 + (i-1)w, iw\}$. For the first part, note that it is sufficient to bound the probability that there does not exist $i$ such that $J_i$ intersects with at three of more of the $t$ intervals in $\mathcal{I}$. But, the probability that a particular $J_i$ intersects with three of more of these is at most $\binom{t}{3}(2w/n)^3 \leq (2tw/n)^3$. Hence the expected number of $i \in [n/w]$ such that $J_i$ intersects with three of more of the $t$ intervals is at most

$$(n/w)(2tw/n)^3 = 8w^2 t^3/n^2 = 8c_1^2 \ .$$

By Markov's inequality the probability there is a $J_i$ that intersects with three of more intervals is at most $8c_1^2$. For the second part, we consider the intervals $J_i$ that overlap with two sets from $\mathcal{I}$. The expected number of such intervals is at most

$$(n/w)\binom{t}{2}(2w/n)^2 \leq 4t^2 w/n = c_1 n^{1/(2k)} \ .$$

Hence, by Markov's inequality, the second event occurs with probability at most $c_1$. □

**Lemma 2.** *Consider a random subset $S \subset [n]$ of size $n^{1/k}$. For sufficiently small constant $c_2$, with probability at least $99/100$, for each $i, j \in S$, $|j - i| \geq c_2 n^{1-2/k}$.*

Lemma 2 follows from an elementary "birthday paradox" analysis. We are now ready to prove our main result.

**Theorem 1.** *Estimating $F_k$ up to a factor 2 in the random-order data-stream model with probability at least $9/10$ requires $\Omega(n^{1-2.5/k}/\log n)$ bits of space.*

*Proof.* Let $\{S_1, \ldots, S_t\}$ be an instance of $t$-party set-disjointness where $|S_i| = c_1 n^{1-3/(2k)} =: w$, $N = c_1 n^{1-1/(2k)}$, and $t = 100 n^{1/k}$. Consider $t$ players where the $i$-th player knows $S_i$. Let $\mathcal{A}$ be a $s$-space, single-pass, data-stream algorithm that returns a 2-approximation for $F_k$ of a $O(n)$-length stream with probability $99/100$ on the assumption that the order of the stream elements is chosen uniformly at random from the set of all orderings.

The players use $\mathcal{A}$ to solve the instance of $t$-party set-disjointness as follows. Using public randomness the players pick:

(1) Sets $I_1 = [a_1, b_1], \ldots, I_t = [a_t, b_t]$ from $\text{Cycle}_{n,w}$ (without loss of generality $b_i \leq b_j$ if $i \leq j$).
(2) A permutation $\sigma$ of $[2n]$.
(3) A length $n^{2/k}/c_2$ random binary string $r$.

If $b_1 < w$ or there exists some $j \in [n]$ that appears in three of the intervals, the protocol terminates with failure. Note that the probability of this event is $(w-1)/n + 1/100 \leq 2/100$ by Lemma 1.

Given the sets $I_1, \ldots, I_t$ we define the intervals

$$A_i = \begin{cases} [a_{i+1}, b_i] & \text{if } b_i > a_{i+1} \\ [b_i + 1, a_{i+1} - 1] & \text{if } b_i \leq a_{i+1} \end{cases}$$

where $b_0 = 0$ and $a_{t+1} = n+1$. We say $A_i$ is a *doubled interval* if $A_i = I_i \cap I_{i+1}$ and call it an *easy interval* otherwise. Let $B_i = I_i \setminus (I_{i-1} \cup I_{i+1})$. Then the $A_i$'s and $B_i$'s are disjoint and,

$$[n] = A_0 \cup B_1 \cup A_1 \cup \ldots B_t \cup A_t .$$

Also consider a partitioning of $[n]$ into $n^{2/k}/c_2$ intervals $C_i$ of length $w_2 = c_2 n^{1-2/k}$, $[n] = \cup C_i$

Player $i$ constructs a string $s_i$ consisting of the elements from $S_i$ in a random order, with $\sigma$ applied to each. The $j$-th entry of the constructed stream is determined by

(1) Player $i$ if $j \in A_{i-1}$ where $A_{i-1}$ is an easy interval and set to be $\sigma(n+j)$
(2) Player $i$ if $j \in B_i$. The element is set to $\sigma(n+j)$ with probability $1/2$, and to $s_\ell^i$ otherwise, where $j$ is the $\ell$-th element of $I_i$.
(3) Player $i$ if $j \in A_{i-1} \cup C_m$ and $r_m = 0$ where $A_i$ is a doubled interval. The element is set to $s_\ell^i$ where $j$ is the $\ell$-th element of $I_i$.
(4) Player $i-1$ if $j \in A_{i-1} \cup C_m$ and $r_m = 1$ where $A_i$ is a doubled interval. The element is set to $s_\ell^{i-1}$ where $j$ is the $\ell$-th element of $I_{i-1}$.

By appealing to Lemma 1, we note that the players can simulate an algorithm on this stream using $O(n^{1/k} + n^{1/(2k)} w/w_2) = O(n^{1/k})$ messages with high probability at least $99/100$. The size of each message is at most $s$.

Hence the space use of the algorithm must be at least $\Omega(n^{1-2.5/k}/\log n)$. With probability at least $99/100$, the multiplicity of the most frequent element of the stream is greater than $(2n)^{1/k}$. Hence, in the case that there exists $j \in [n]$ such that $j \in S_i$ for all $i \in [t]$, $F_k \geq 2n$ and otherwise $F_k \leq n$. Therefore a 2-approximation of the $F_k$ solves the instance of $t$-party set-disjointness. It remains to show that the ordering of the stream is near random so that we may assume that $\mathcal{A}$ returns a 2-approximation of $F_k$ as required. This follows because the location of the multiply occurring element (if one exists) were chosen by picking $t$ random positions and then deleting each occurrence independently with probability $1/2$ (by Lemma 2, we may condition on the fact that no two elements occur within $w_2$ of each other). Hence the probability that the protocol succeeds is at least $99/100 - 2/100 - 1/100 - 100 = 19/20$. □

## References

[1] N. Alon, Y. Matias, and M. Szegedy. The space complexity of approximating the frequency moments. *Journal of Computer and System Sciences*, 58(1):137–147, 1999.
[2] L. Bhuvanagiri, S. Ganguly, D. Kesh, and C. Saha. Simpler algorithm for estimating frequency moments of data streams. In *ACM-SIAM Symposium on Discrete Algorithms*, pages 708–713, 2006.

[3] A. Chakrabarti, G. Cormode, and A. McGregor. Robust lower bounds for communication and stream computation. *ACM Symposium on Theory of Computing*, 2008.

[4] A. Chakrabarti, T. Jayram, and M. Pătraşcu. Tight lower bounds for selection in randomly ordered streams. In *ACM-SIAM Symposium on Discrete Algorithms*, 2008.

[5] A. Chakrabarti, S. Khot, and X. Sun. Near-optimal lower bounds on the multi-party communication complexity of set disjointness. In *IEEE Conference on Computational Complexity*, pages 107–117, 2003.

[6] E. D. Demaine, A. López-Ortiz, and J. I. Munro. Frequency estimation of internet packet streams with limited space. In *European Symposium on Algorithms*, pages 348–360, 2002.

[7] S. Guha and A. McGregor. Space-efficient sampling. In *AISTATS*, pages 169–176, 2007.

[8] S. Guha and A. McGregor. Stream order and order statistics: Quantile estimation in random-order streams. *SIAM Journal of Computing*, 2008.

[9] P. Indyk and D. P. Woodruff. Optimal approximations of the frequency moments of data streams. In *ACM Symposium on Theory of Computing*, pages 202–208, 2005.

[10] A. McGregor. *Processing Data Streams*. PhD thesis, University of Pennsylvania, 2007.

[11] J. I. Munro and M. Paterson. Selection and sorting with limited storage. *Theor. Comput. Sci.*, 12:315–323, 1980.

[12] D. P. Woodruff. Optimal space lower bounds for all frequency moments. In *ACM-SIAM Symposium on Discrete Algorithms*, pages 167–175, 2004.